\documentclass{article}

\usepackage[english]{babel}

\usepackage[letterpaper,top=2cm,bottom=2cm,left=3cm,right=3cm,marginparwidth=1.75cm]{geometry}

\usepackage{amssymb}
\usepackage{textcomp}
\usepackage{amsmath}
\usepackage{algorithm}
\usepackage{algpseudocode}

\usepackage{graphicx}
\usepackage[colorlinks=true, allcolors=blue]{hyperref}

\newtheorem{theorem}{Theorem}
\newtheorem{property}{Property}
\newtheorem{corollary}{Corollary}

\newcommand{\ket}[1]{|#1\rangle}
\newcommand{\Beginproof}{{\em Proof.}  }
\newcommand{\Endproof}{\hfill$\Box$\\}

\title{Quantum Algorithm for the Longest Trail Problem}
\author{Kamil Khadiev and Ruslan Kapralov}
\date{Kazan Federal University, Kazan, Russia\\\small \texttt{email: kamilhadi@gmail.com}}
\begin{document}
\maketitle

\begin{abstract}
We present the quantum algorithm for the Longest Trail Problem. The problem is to search the longest edge-simple path for a  graph with $n$ vertexes and $m$ edges. Here edge-simple means no edge occurs in the path twice, but vertexes can occur several times. The running time of our algorithm is $O^*(1.728^m)$.
\end{abstract}


\section{Introduction} Quantum computing
\cite{nc2010,a2017,aazksw2019part1} is one of the hot topics in computer science of the last decades.
There are many problems where quantum algorithms outperform the best-known classical algorithms. Some of them can be founded here \cite{dw2001,quantumzoo}. 
Problems for graphs are examples of such problems \cite{ks2019,kks2019,abikpv2019,dhhm2004}. One of the most important performance metrics in this regard is \emph{query complexity}; and we investigate problems using this metric for complexity. 

In this paper, we consider the Longest Trail Problem (LTP). The problem is the following one. Let us consider a graph with $n$ vertexes and $m$ edges.
The problem is to search the {\em longest edge-simple} path. Here {\em edge-simple} means no edge occurs in the path twice, but vertexes can occur several times. The {\em longest} means the path has the maximal possible number of edges.  

The problem is strongly related to the longest path problem (LPP) that is searching the longest {\em vertex-simple} path. Here {\em vertex-simple} means no vertex occurs in the path twice.

There are many practical applications of these problems, for example, \cite{abcc2007,siks2006}.

Both problems are NP-hard \cite{l2001CombOpt}. The NP-hardness of LTP problem was discussed in \cite{ltp2014}.

The simple classical solution for the problem can be a brute force algorithm that checks all possible paths and searching the required one. Such solution works in $O(m!)=O(m^{m})$ running time. This solution can be used as a base of a quantum algorithm because the classical algorithm solves a search problem. Therefore, we can use  Grover Search algorithm \cite{g96,bbht98} and obtain a quantum algorithm that works in $O(\sqrt{m!})=O(m^{0.5m})$. At the same time, there is a better classical algorithm that is based on the Dynamic programming approach \cite{b62,hk62}. This classical algorithm for the LTP problem works in $O^*(2^m)$ running time, where $O^*$ hides polylog factors. The algorithm is not a simple search algorithm. That is why we cannot directly use the Grover Search algorithm for quantum speed-up, and we cannot obtain a complexity $O^*(1.4^m)$ using this way. 
We present a quantum algorithm that works in $O^*(1.728^m)$ running time. The algorithm is based on Grover Search algorithm \cite{g96,bbht98}, quantum minimum finding algorithm \cite{dh96,dhhm2004} and quantum ideas for dynamic programming on Boolean cube \cite{abikpv2019}.

The structure of the paper is the following. Section \ref{sec:prelims} contains preliminaries.  Then, we discuss algorithm in Section \ref{sec:algo}.

\section{Preliminaries}\label{sec:prelims}

\subsection{The Longest Trail Problem} Let $G=(V,E)$ be an unweighted, underacted graph, where $V$ is a set of vertexes, and $E$ is a set of edges. Let $m=|E|$ be the number of edges and $n=|V|$ be the number of vertexes.

Let a sequence of edges $P=(e_{i_1},\dots,e_{i_{\ell}})$ be a path if each sequentially pair of edges $e_{i_{j}}$ and $e_{i_{j+1}}$ has common vertex, for $j\in\{1,\dots,\ell-1\}$. A path is edge-simple if the sequence has no duplicates i.e., for any $j\neq j'$ we have $e_{i_j}\neq e_{i_{j'}}$. Let $|P|=\ell$ be the length of a path $P$. Let ${\cal P}(G)$ be the set of all possible paths for a graph $G$.

The problem is to the longest path i.e., any path $P_{long}$ such that $|P_{long}|=max\{|P|:P\in{\cal P}(G) \}$.

\subsection{Quantum Query Model}
We use the standard form of the quantum query model. 
Let $f:D\rightarrow \{0,1\},D\subseteq \{0,1\}^N$ be an $N$ variable function. An input for the function is $x=(x_1,\dots,x_N)\in D$ where $x_i\in\{0,1\}$ for $i\in\{1,\dots,N\}$.

We are given oracle access to the input $x$, i.e. it is realized by a specific unitary transformation usually defined as $\ket{i}\ket{z}\ket{w}\rightarrow \ket{i}\ket{z+x_i\pmod{2}}\ket{w}$ where the $\ket{i}$ register indicates the index of the variable we are querying, $\ket{z}$ is the output register, and $\ket{w}$ is some auxiliary work-space. It can be interpreted as a sequence of control-not transformations such that we apply inversion operation (X-gate) to the second register that contains $\ket{z}$ in a case of the first register equals $i$ and the variable $x_i=1$. We interpret the oracle access transformation as $N$ such controlled transformations for each $i\in\{1,\dots,N\}$.   

An algorithm in the query model consists of alternating applications of arbitrary unitaries independent of the input and the query unitary, and a measurement in the end. The smallest number of queries for an algorithm that outputs $f(x)$ with a probability that is at least $\frac{2}{3}$ on all $x$ is called the quantum query complexity of the function $f$ and is denoted by $Q(f)$. We refer the readers to \cite{nc2010,a2017,aazksw2019part1} for more details on quantum computing. 

In this paper's quantum algorithms, we refer to the quantum query complexity as the quantum running time. We use modifications of Grover's search algorithm \cite{g96,bbht98} as quantum subroutines. For these subroutines, time complexity is more than query complexity for additional log factor.

\section{Algorithm}\label{sec:algo}

We discuss our algorithm in this section.
Let us consider a function $L:2^E\times E\times E\to \mathbb{R}$ where $2^E$ is the set of all subsets of $E$. The function $L$ is such that $L(S,v,u)$ is the length of the longest path that uses only edges from the set $S$, starts from the edge $v$, and finishes in the edge $u$.

Let the function $F:2^E\times E\times E\to E^*$ be such that $F(S,v,u)$ is the longest path that uses only edges from the set $S$, starts from the edge $v$, and finishes in the edge $u$.

It is easy to see that $L(\{v\},v,v)=1$ and $F(\{v\},v,v)=(v)$ for any $v\in E$ because the set has only one edge and it is the only path in the set.

Another property of these functions is

\begin{property} Suppose $S\in 2^E, v,u\in E$, an integer $k\leq |S|$. The function $L$ is such that
\[L(S,v,u)=\max\limits_{S'\subset S,|S'|=k,y\in S'}\left(L(S',v,y)+L((S\backslash S') \cup \{y\},y,u)\right)\]
and $F(S,u,v)$ is the path that is concatenation of corresponding paths.
\end{property}
\Beginproof
Let $P^1=F(S',v,y)$ and $P^2=F((S\backslash S') \cup \{y\},y,u)$. The path $P=P^1\circ P^2$ belongs to $S'$, starts from $v$ and finishes in $u$, where $\circ$ means concatenation of paths excluding the duplication of common edge $y$. Because of definition of $L$, we have $L(S,v,u)\geq |P|$.

Assume that there is $T=(e_1,\dots,e_\ell)$ such that $\ell=|T|=L(S,v,u)$ and $|T|>|P|$. Let us select $S''$ such that $|S''|=k, S''\subset S$ and there is $j<|T|$ such that $R^1=e_1,\dots,e_j\in S''$ and $R^2=e_j,e_{j+1},\dots,e_\ell\not\in S''\backslash\{e_j\}$. Then $|R^1|\leq |P^1|$ and $|R^2|\leq |P^2|$ by definition of $F$ and $L$. Therefore, $|R|=|R^1|+|R^2|-1\leq |P^1|+|P^2|-1=|P|$. We obtain a contradiction with assumption.
\Endproof

As a corollary we obtain the following result:
\begin{corollary}\label{cor:one-edge}
Suppose $S\in 2^E, v,u\in E$, ${\cal I}(u)$ is the set of all edges that has common vertex with $u$. The function $L$ is such that
\[L(S,v,u)=\max\limits_{y\in S\backslash\{u\}, y\in {\cal I}(u)}\left(L( S\backslash\{u\},v,y)+1\right)\]
and $F(S,u,v)$ is the path that is the corresponding path.
\end{corollary}
Using this idea, we construct the following algorithm.

{\bf Step 1}. Let $\alpha=0.055$. We classically compute $L(s,v,u)$ and $F(S,v,u)$ for $|S|=(1-\alpha)\frac{m}{4}$ and $v,u\in E$ 

{\bf Step 2}. Let $E_4\subset E$ be such that $|E_4|=\frac{m}{4}$. Then, we have 

\[L(E_4,u,v)=\max\limits_{E_{\alpha}\subset E_4,|E_{\alpha}|=(1-\alpha)m/4,y\in E_{\alpha}}\left(L(E_{\alpha},v,y)+L((E_4\backslash E_{\alpha}) \cup \{y\},y,u)\right).\]

Let $E_2\subset E$ be such that $|E_2|=\frac{m}{2}$. Then, we have

\[L(E_2,u,v)=\max\limits_{E_4\subset E_2,|E_4|=m/4,y\in E_4}\left(L(E_4,v,y)+L((E_2\backslash E_4) \cup \{y\},y,u)\right).\]

Finally,
\[L(E,u,v)=\max\limits_{E_2\subset E,|E_2|=m/2,y\in E_2}\left(L(E_2,v,y)+L((E\backslash E_2) \cup \{y\},y,u)\right).\]

We can compute $L(E,u,v)$ and corresponding $F(E,u,v)$ using three nested procedures for maximum finding. As such procedure, we use  Durr-Hoyer \cite{dh96,dhhm2004} quantum minimum finding algorithm.

Note that the error probability for the Durr-Hoyer algorithm is at most $0.1$. So, we use the standard boosting technique to decrease the total error probability to constant by $O(m)$ repetition of the maximum finding algorithm in each level.

Let us present the implementation of Step 1. Assume that ${\cal I}(u)$ is the sequence of edges that have a common vertex with the edge $u$.
Let us present a recursive function $\textsc{GetLen}(S,v,u)$ for $S\in 2^E,u,v\in E$ with cashing that is Dynamic Programming approach in fact. The function is based on Corollary \ref{cor:one-edge}.

\begin{algorithm}
\caption{$\textsc{GetLen}(S,v,u)$.}
\begin{algorithmic}
\If{$v=u$ and $S=\{v\}$}\Comment{Initialization}
\State $L(\{v\},v,v)\gets 1$
\State $F(\{v\},v,v)\gets (v)$
\EndIf
\If {$L(S,v,u)$ is not computed}
\State $len\gets -1$
\State $path\gets ()$
\For{$y \in {\cal I}(u)$}
\If{$y\in S\backslash\{u\}$ and $\textsc{GetLen}( S\backslash\{u\},v,y)+1>len$}
\State $len\gets L( S\backslash\{u\},v,y)+1$
\State $path\gets F( S\backslash\{u\},v,y)\cup u$
\EndIf
\EndFor
\State $L(S,v,u)\gets len$
\State $F(S,v,u)\gets path$
\EndIf
\State \Return{$L(S,v,u)$}
\end{algorithmic}
\end{algorithm}

\begin{algorithm}
\caption{$\textsc{Step1}$.}
\begin{algorithmic}
\For{$S \in 2^E$ such that $|S|=(1-\alpha)\frac{m}{4}$}
\For{$v\in E$}
\For{$u\in E$}
\If{$v \in S$ and $u\in S$}
\State $\textsc{GetLen}(S,v,u)$\Comment{We are computing $L(S,v,u)$ and $F(S,v,u)$ but we are not needing this results at the moment. We need it for Step 2.}
\EndIf
\EndFor
\EndFor
\EndFor
\end{algorithmic}
\end{algorithm}

Let $\textsc{QMax}((x_1,\dots,x_N))$ be the implementation of the quantum maximum finding algorithm \cite{dh96} for a sequence $x_1,\dots,x_N$.

The most nested quantum maximum finding algorithm for some $E_4\subset E, |E_4|=\frac{m}{4}$ and $u,v\in E$ is \[\textsc{QMax}((L(E_{\alpha},v,y)+L((E_4\backslash E_{\alpha}) \cup \{y\},y,u):E_{\alpha}\subset E_4,|E_{\alpha}|=(1-\alpha)\frac{m}{4},y\in E_{\alpha}))\]

The middle quantum maximum finding algorithm for some $E_2\subset E, |E_2|=\frac{m}{2}$ and $u,v\in E$ is 

\[\textsc{QMax}((L(E_{4},v,y)+L((E_2\backslash E_{4}) \cup \{y\},y,u):E_4\subset E_2,|E_4|=n/4,y\in E_4))\]

Note that $|E_4|=m/4$ and $|E_2\backslash E_{4}|=m/4$. We use the invocation of $\textsc{QMax}$ (the most nested quantum maximum finding algorithm) instead of $L(E_{4},v,y)$  and $L(E_2\backslash E_{4},y,u)$.

The final quantum maximum finding algorithm for some  $u,v\in E$ is 

\[\textsc{QMax}((L(E_2,v,y)+L((E\backslash E_2) \cup \{y\},y,u):E_2\subset E,|E_2|=m/2,y\in E_2))\]

Note that $|E_2|=m/2$ and $|E\backslash E_2|=m/2$. We use the invocation of $\textsc{QMax}$ (the middle quantum maximum finding algorithm) instead of $L(E_2,v,y)$ and $L((E\backslash E_2) \cup \{y\},y,u)$.

The procedure $\textsc{QMax}$ returns not only the maximal value, but the index of the target element. Therefore, by the ``index'' we can obtain the target paths using $F$ function. So the result path is
$P=P^1\circ P^2$, where $P^1$ is the result path for $L(E_2,v,y)$ and $P^2$ is the result path for $L((E\backslash E_2) \cup \{y\},y,u)$.

$P^1=P^{1,1}\circ P^{1,2}$, where $P^{1,1}$ is the result path for $L(E_{4},v,y)$ and $P^{1,2}$ is the result path for $L((E_2\backslash E_{4}) \cup \{y\},y,u)$. By the same way we can construct $P^2=P^{2,1}\circ P^{2,2}$.

$P^{1,1}=P^{1,1,1}\circ P^{1,1,2}$, where $P^{1,1,1}$ is the result path for $L(E_{\alpha},v,y)$ and $P^{1,1,2}$ is the result path for $L((E_4\backslash E_{\alpha}) \cup \{y\},y,u)$. Note, that this values were precomputed classically, and were stored in $F(E_{\alpha},v,y)$ and $F((E_4\backslash E_{\alpha}) \cup \{y\},y,u)$ respectively.

By the same way we can construct
\[P^{1,2}=P^{1,2,1}\circ P^{1,2,2},\]
\[P^{2,1}=P^{2,1,1}\circ P^{2,1,2},\]
\[P^{2,2}=P^{2,2,1}\circ P^{2,2,2}.\]

The final Path is
\[P=P^1\circ P^2=(P^{1,1}\circ P^{1,2})\circ(P^{2,1}\circ P^{2,2})=\]
\[\Big((P^{1,1,1}\circ P^{1,1,2})\circ (P^{1,2,1}\circ P^{1,2,2})\Big)\circ\Big((P^{2,1,1}\circ P^{2,1,2})\circ (P^{2,2,1}\circ P^{2,2,2})\Big)\]

Let us present the final algorithm as Algorithm \ref{alg:main}.
\begin{algorithm}
\caption{Algorithm for LTP.}\label{alg:main}
\begin{algorithmic}
\State $\textsc{Step1}()$
\State $len\gets -1$
\State $path \gets ()$
\For{$v\in E$}
\For{$u\in E$}
\State $currentLen\gets\textsc{QMax}((L(E_2,v,y)+L((E\backslash E_2) \cup \{y\},y,u):E_2\subset E,|E_2|=m/2,y\in E_2))$
\If{$len<currentLen$}
\State $len\gets currentLen$
\State $path\gets \Big((P^{1,1,1}\circ P^{1,1,2})\circ (P^{1,2,1}\circ P^{1,2,2})\Big)\circ\Big((P^{2,1,1}\circ P^{2,1,2})\circ (P^{2,2,1}\circ P^{2,2,2})\Big)$
\EndIf
\EndFor
\EndFor
\State \Return{path}
\end{algorithmic}
\end{algorithm}

The complexity of the algorithm is presented in the next theorem.
\begin{theorem}
 Algorithm \ref{alg:main} solves LTP with $O^*\left(1.728^m\right)$ running time and constant bounded error.
\end{theorem}
\Beginproof
The correctness of the algorithm follows from the above discussion. Let us present an analysis of running time.

Complexity of Step 1 (Classical preprocessing) is \[O^*\left( \binom{m}{(1-\alpha) \frac{m}{4}}\right)=O^*(1.728^m).\]

Complexity of Step 2 (Quantum part) is complexity of three nested Durr-Hoyer maximum finding algorithms. Due to  \cite{dh96,g96,dhhm2004,aazksw2019part1}, the complexity is

\[O^*\left(\sqrt{\binom{m}{m/2}}\cdot\sqrt{\binom{m/2}{m/4}}\cdot\sqrt{\binom{m/4}{\alpha m/4}}\right)=O^*(1.728^m).\]

We invoke Step 1 and Step 2 sequentially. Therefore the total complexity is the sum of complexities for these steps. So, the total complexity is $O^*(1.728^m)$.

Only Step 2 has an error probability. The most nested invocation of the Durr-Hoyer algorithm has an error probability $0.1$. Let us repeat it $2m$ times and choose the maximal value among all invocations. The algorithm has an error only if all invocations have an error. Therefore, the error probability is $0.1^{2m}=100^{-m}$.

Let us consider the middle Durr-Hoyer algorithm's invocation. The probability of success is the probability of correctness of maximum finding and the probability of input correctness, i.e., the correctness of all the nested Durr-Hoyer algorithm's invocations. It is
\[0.9\cdot  (1-100^{-m})^{\gamma}\mbox{, where }\gamma= \binom{m/2}{m/4}\]
\[\geq 0.8\mbox{, for enough big }m.\] 
So, the error probability is at most $0.2$.

 Let us repeat the middle Durr-Hoyer algorithm $2m$ times and choose the maximal value among all invocations. Similar to the previous analysis, the error probability is $0.2^{2m}=25^{-m}$.
 
 Therefore, the total success probability that is the final Durr-Hoyer algorithm's success probability is the following one.
 \[0.9\cdot  (1-25^{-m})^{\beta}\mbox{, where }\beta= \binom{m}{m/2}\]
 \[>0.8\mbox{, for enough big }m.\]
 Therefore, the total error probability is at most $0.2$.
\Endproof


\subsection*{Acknowledgments}
The research is funded by the subsidy allocated to Kazan Federal University for the state assignment in the sphere of scientific activities, project No. 0671-2020-0065.


\bibliographystyle{plain}
\bibliography{tcs}
\end{document}